\documentclass[showpacs, floatfix,twocolumn]{revtex4}
\usepackage{amsmath}
\usepackage{color}
\usepackage{amssymb}

\usepackage{graphicx}

\begin{document}

\title{Phase diagram of UCoGe}

\author{V.P.Mineev}
\affiliation{$^1$
 Commissariat a l'Energie Atomique, UGA, INAC-FELIQS, 38000 Grenoble, France}

\begin{abstract}
The temperature-pressure phase diagram of ferromagnetic superconductor UCoGe includes four phase transitions. They are between the paramagnetic and the ferromagnetic states with the subsequent transition in the superconducting ferromagnetic state and between the normal and the superconducting states after which has to occur the transition to the superconducting ferromagnetic state. Here we have developed
the Landau theory description of the phase diagram and established the specific ordering arising at each type of transition.

The phase transitions to the ferromagnetic superconducting state are inevitably accompanied by the emergency of screening currents. The corresponding magnetostatics considerations  allow to establish the significant difference between   the transition from the ferromagnetic to the ferromagnetic superconducting state and the transition from the superconducting to the ferromagnetic superconducting state.

\end{abstract}

\date{\today}
\maketitle
\section{Introduction}

The superconductivity in the uranium ferromagnetic compounds UGe$_2$ and URhGe discovered more than decade ago  \cite{Saxena01,Aoki01} and more recently  
in the related compound UCoGe  \cite{Huy07} is still the subject of quite active investigations (see  recent experimental \cite{Aoki14} and theoretical \cite{Mineev16} reviews and references therein). The existence of superconducting state  at temperatures far below the Curie temperature and very high upper critical field  in these materials 
do not leave doubts that here we deal with the  triplet superconductivity  like it is  in the superfluid $^3$He. This is also confirmed by the measurements of the Knight shift on nucleus of $^{59}$Co which is proved unchanged in the superconducting state \cite{Hattori13}.

One of many peculiar properties of UCoGe is that the ferromagnetism in this compound is suppressed by pressure  whereas the superconductivity
arising at small pressures inside of the ferromagnetic state continues to exist at high pressures in the paramagnetic state.The pressure-temperature phase diagram shown  in Fig.1, has been established first in the paper \cite{Slooten09}  and then confirmed  in many subsequent studies (see f.i. the last one \cite{Bastien16}). The phase transition from the paramagnetic to the ferromagnetic state  and the following to it the  phase transition to the ferromagnetic superconducting state at low pressures and the phase transition from the normal to the superconducting  state at high pressures are firmly registered. While the phase transition from the  superconducting to the ferromagnetic superconducting state shown on the Fig.1 by the dashed line is still not confirmed experimentally. 

A theoretical   phase diagram description has been proposed recently
   by Cheung and Raghu \cite{Raghu16}. Making use the numerical calculations applied to  the minimal Landau model of neutral ferromagnetic superfluid state  with one component order parameter for each   spin up-up and spin down-down  Cooper pair states they were able to reproduce the general structure of the UCoGe phase diagram and to predict a first order phase transition near the boundary between the normal phase and the ferromagnetic superconducting phase.

Here I reconsider the same problem making use the
analytical calculations applied to the same minimal model for neutral ferromagnetic or nonmagnetic superfluid states. The results of Ref.9 were confirmed.

A phase transition of normal metallic  to the superconducting state has its own specific properties different from the properties of a phase transition in the neutral Fermi liquid to the superfluid state. So, in the last part of the paper 
 I will discuss the significant   difference   between  the two  transitions, namely, between the phase transition from 
the  ferromagnetic normal  state to the ferromagnetic superconducting state and the  transition from the  superconducting state to the ferromagnetic superconducting state. This difference arises due to the essentially different
 screening of magnetic moment  at these two transitions.
 In the latter case the screening is complete and instead of a bulk  phase transition  there is the gradual formation of the Meissner state
as it occurs  in a superconductor of the second kind under external magnetic field smaller than $H_{c1}$.

\section{Model}

The triplet-pairing superconducting state order parameter is given by the complex spin-vector \cite{Book} 
\begin{eqnarray}
{\bf d}({\bf k},{\bf r})=~~~~~~~~~~~~~~~~~~~~~~~~~~~\\
\frac{1}{2}
\left[-\Delta^{\uparrow}¥({\bf k},{\bf r})(\hat{x}+i\hat{y})+
\Delta^{\downarrow}¥({\bf k},{\bf r})(\hat{x}-i\hat{y})\right]+\Delta^{0}¥({\bf k},{\bf r})\hat z,\nonumber
\label{e14}
\end{eqnarray}
where $\Delta_{\uparrow}¥({\bf k},{\bf r})$,  $\Delta_{\downarrow}¥({\bf r},{\bf k},{\bf r})$, $ \Delta_{0}¥({\bf k},{\bf r}) $ are the amplitudes of spin-up, spin-down and zero-spin  of superconducting order parameter depending on the Cooper pair centre of gravity coordinate ${\bf r}$ and the momentum ${\bf k }$ of pairing electrons.  In the tetragonal ferromagnets with easy axis along $\hat z$ direction
there are only two superconducting states  A and B  with different critical temperature  \cite{Mineev02}.  The general form of the order parameter for the A-state in a two-band spin-up, spin-down superconducting ferromagnet  
\begin{eqnarray}
&\Delta_A^\uparrow({\bf k},{\bf r})=\hat k_x\eta_x^\uparrow({\bf r})+i\hat k_y\eta_y^\uparrow({\bf r}),\nonumber\\
&\Delta_A^\downarrow({\bf k},{\bf r})=\hat k_x\eta_x^\downarrow({\bf r})+i\hat k_y\eta_y^\downarrow({\bf r}),\label{A}\\
&\Delta_A^0({\bf k},{\bf r})=\hat k_z\eta_z^0({\bf r})\nonumber
\end{eqnarray}
depends from the five complex amplitudes $\eta_x^\uparrow,~ \eta_y^\uparrow,~
\eta_x^\downarrow,~
\eta_y^\downarrow,~
\eta_z^0$, which   obey to coupled Ginzburg-Landau equations, derived in the linear approximation in the papers \cite{Mineev14,Mineev16}.
$\hat k_i=k_i/|{\bf  k}|,~~~i=x,y,z$ are the projections of the unit $\hat{\bf k}$ vector on the coordinate axis.

 The order parameter of the paramagnetic superconducting state \cite{Mineev09,Mineev16} in a orthorhombic metal looks like the order parameter of superfluid $^3$He-B phase \cite{Book}
\begin{eqnarray}
&\Delta^\uparrow({\bf k},{\bf r})=-\hat k_x\eta_x({\bf r})+i\hat k_y\eta_y({\bf r}),\nonumber\\
&\Delta^\downarrow({\bf k},{\bf r})=\hat k_x\eta_x({\bf r})+i\hat k_y\eta_y({\bf r}),\label{B}\\
&\Delta_A^0({\bf k},{\bf r})=\hat k_z\eta_z({\bf r}).\nonumber
\end{eqnarray}

To avoid excessive difficulties the authors of \cite{Raghu16} considered the minimal model  for the  superconducting ferromagnetic state with
the order parameter
\begin{eqnarray}
&\Delta^\uparrow({\bf k},{\bf r})=\hat k_x\eta_\uparrow({\bf r})
,\nonumber\\
&\Delta^\downarrow({\bf k},{\bf r})=\hat k_x\eta_\downarrow({\bf r}).
\label{A'}
\end{eqnarray}
Corresponding simplest order parameter for  the paramagnetic superconducting state  looks like the order parameter for the discovered recently polar state of superfluid $^3$He  \cite{Dmitriev}
\begin{eqnarray}
&\Delta^\uparrow({\bf k},{\bf r})=-\hat k_x\eta({\bf r})
,\nonumber\\
&\Delta^\downarrow({\bf k},{\bf r})=\hat k_x\eta({\bf r}).
\label{B'}
\end{eqnarray}

In neglect of interactions of electron charges with magnetic field created by the magnetization  
one can write following Ref.9 the gradient independent Landau free energy density as
\begin{widetext}
\begin{equation}
F=\alpha M^2+\beta M^4 + \alpha_1(|\eta_\uparrow|^2+|\eta_\downarrow|^2)+\gamma_1M(|\eta_\uparrow|^2-|\eta_\downarrow|^2)+
\gamma_2(\eta_\uparrow\eta_\downarrow^\star+\eta_\uparrow^\star\eta_\downarrow)+B(|\eta_\uparrow|^2+|\eta_\downarrow|^2)^2+
C(|\eta_\uparrow|^2-|\eta_\downarrow|^2)^2,
\label{fe}
\end{equation}
\end{widetext}
where $M$ is the density of magnetic moment component along the easy axis,
\begin{equation}
\alpha=\alpha_0(T-T_c),~~~~~\alpha_1=\alpha_{10}(T-T_{sc0}),
\end{equation}
$T_c( P )$ is the pressure dependent Curie temperature and $T_{sc0}( P )$ is the formal  critical temperature of superconducting transition in the single band (say just spin-up) case. 
 The phenomenological treatment does not allow to fix the  pressure dependences of these critical  temperatures and the other coefficients in Eq.(\ref{fe}).
In what follows we shall assume that the pressure dependences $T_c( P )$ and $T_{sc0}( P )$
qualitatively correspond to the phase diagram with the intersection of the phase transition lines shown in 
Fig.1.

One can note that the symmetry allows also  the following interaction $i\gamma_3M(\eta_\uparrow\eta_\downarrow^\star-\eta_\uparrow^\star\eta_\downarrow)$  between the superconducting and magnetic order parameters \cite{Mineev09}, but the general enough microscopic calculations \cite{Mineev14,Mineev16} do not confirm the existence of this term.

In general, the free energy fourth order terms actually have the form different from  $B(|\eta_\uparrow|^2+|\eta_\downarrow|^2)^2+
C(|\eta_\uparrow|^2-|\eta_\downarrow|^2)^2$ used in Ref.9.  If the normal state Green functions are diagonal in the band indices 
(in our case  they are spin-up and spin-down indices ) the Wick's decoupling does not produce any mixing terms between the band order parameters (see f.i. \cite{Dao}). This case the fourth order terms  in respect of the superconducting order parameters are
\begin{equation}
\beta_1(|\eta_\uparrow|^4+|\eta_\downarrow|^4)+\tilde\beta_1M(|\eta_\uparrow|^4-|\eta_\downarrow|^4).
\end{equation}
If in the normal state there is the band mixing interaction  this leads  to emergence of the additional terms
\begin{widetext}
\begin{equation}
\beta_2|\eta_\uparrow|^2|\eta_\downarrow|^2+\beta_3[(\eta_\uparrow\eta_\downarrow^\star)^2+(\eta_\uparrow^\star\eta_\downarrow)^2]
+\beta_4(|\eta_\uparrow|^2+|\eta_\downarrow|^2)(\eta_\uparrow\eta_\downarrow^\star+\eta_\uparrow^\star\eta_\downarrow)+\tilde\beta_4M
(|\eta_\uparrow|^2-|\eta_\downarrow|^2)(\eta_\uparrow\eta_\downarrow^\star+\eta_\uparrow^\star\eta_\downarrow).
\end{equation}
\end{widetext}
One can show that the additional fourth order terms do not introduce a qualitative modification in the phase diagram. So, we will work with the same free energy density as in Ref.9
\begin{widetext}
\begin{eqnarray}
F=\alpha M^2+\beta M^4 + \alpha_1(|\eta_\uparrow|^2+|\eta_\downarrow|^2)+\gamma_1M(|\eta_\uparrow|^2-|\eta_\downarrow|^2)+
\gamma_2(\eta_\uparrow\eta_\downarrow^\star+\eta_\uparrow^\star\eta_\downarrow)+
\beta_1(|\eta_\uparrow|^4+|\eta_\downarrow|^4)+
\beta_2|\eta_\uparrow|^2|\eta_\downarrow|^2.
\label{FE}
\end{eqnarray}
\end{widetext}

\section{Phase transitions in neutral Fermi liquid}

At low pressures the system first passes from the paramagnetic to the ferromagnetic state and then
from the ferromagnetic state to the ferromagnetic superconducting state.
We begin with  consideration of these phase transitions and then discuss the   high pressures transitions from the normal  to the superconducting state and
from the superconducting state to the ferromagnetic superconducting state, and also as well as  the transition 
from the  normal to the ferromagnetic superconducting state.

\subsection{Phase transition from the paramagnetic to the ferromagnetic  state}

The second order transition  from paramagnetic to ferromagnetic state  occurs at $T=T_{Curie}( P )$.
Below this temperature the magnetic moment acquires the finite value and a superconducting ordering is absent
\begin{equation}
M^2=(M_0(T))^2=-\frac{\alpha_0(T-T_{c}( P )}{2\beta},~~\eta_\uparrow=\eta_\downarrow=0.
\label{M}
\end{equation}

\subsection{Phase transition from the  ferromagnetic state to  the superconducting ferromagnetic state}

At the subsequent phase transition the superconducting order parameter amplitudes $\eta_\uparrow, \eta_\downarrow$ appear and the magnetic moment acquires a magnitude $M=M_0+m$. Accepting for certainty that coefficient $\gamma_2=-|\gamma_2|$ is negative we see from Eq. (\ref{FE}) that 
the  phase difference between the superconducting order parameters is absent
\begin{equation}
\eta_\uparrow=\eta_1e^{i\varphi},~~~~~ \eta_\downarrow=\eta_2e^{i\varphi}.
\label{eta1eta2}
\end{equation}
Here, $\eta_1$ and $\eta_2$ are the modules of the superconducting order parameters. Thus, one can rewrite the free energy density (\ref{FE}) as
\begin{eqnarray}
F=\alpha M^2+\beta M^4 + \alpha_1(\eta_1^2+\eta_2^2)+\gamma_1M(\eta_1^2-\eta_2^2)\nonumber\\-
2|\gamma_2|\eta_1\eta_2+
\beta_1(\eta_1^4+\eta_2^4)+\beta_2\eta_1^2\eta_2^2.~~~~~~~~~~~~~~
\label{FE'}
\end{eqnarray}

The minimization of the free energy density (\ref{FE'}) in respect $\eta_1, \eta_2$ and $m$  yields  the equations
\begin{widetext}
\begin{eqnarray}
&\alpha_1\eta_1+\gamma_1(M_0+m)\eta_1-|\gamma_2|\eta_2
+2\beta_1\eta_1^3
+\beta_2\eta_1\eta_2^2=0,~
\label{eta1}
\\
&\alpha_1\eta_2-\gamma_1(M_0+m)\eta_2-|\gamma_2|\eta_1
+2\beta_1\eta_2^3+
\beta_2\eta_1^2\eta_2
=0,~\label{eta2}
\\
&2\alpha m +12\beta M_0^2m +12\beta M_0m^2 +4\beta m^3+ \gamma_1(\eta_1^2-\eta_2^2)
=0.~~
\label{mo}
\end{eqnarray}
\end{widetext}
Here, we have taken  into account that $M_0$
is the  minimum of free energy at $\eta_1= \eta_2=0$ and omitted the fourth order terms.
The corresponding linear equations for $\eta_1$, $\eta_2$ 
\begin{eqnarray}
(\alpha_1+\gamma_1M_0)\eta_1-|\gamma_2|\eta_2=0,\\
-|\gamma_2|\eta_1+(\alpha_1-\gamma_1M_0)\eta_2=0
\end{eqnarray}
are not coupled with linear equation for $m$.
Equating the determinant of this system to zero and taking into account Eq.(\ref{M}) we obtain the equation 
\begin{equation}
T_{sc}=T_{sc0}+ \frac
{\sqrt{(\gamma_1(M_0(T_{sc}) )^2+\gamma_2^2}}
{\alpha_{10}}
\label{al}
\end{equation}
for the temperature $T_{sc}$ of transition to the superconducting ferromagnetic state.  We shall not write the explicit formula for  $T_{sc}$
in view of its cumbersome shape. Let us only note  that according to this equation the pressure decrease of the Curie temperature $T_c( P )$
causes the increase of the superconducting transition temperature $T_{sc}( P )$, although   this is not  the only reason for the $T_{sc}( P )$ pressure dependence. 

The linear equation in respect of $m$ gives  
\begin{equation}
m\cong -\frac{\gamma_1(\eta_1^2-\eta_2^2)}{8\beta M_0^2}.
\label{m}
\end{equation}
So, $m$ is proved to be of  the next order of smallness  in comparison with $\eta_1\propto \eta_2\propto\sqrt{T_{sc}-T}$.
Substitution Eq.(\ref{m}) to the Eqs. (\ref{eta1}) and (\ref{eta2}) gives the  equations of the  third order in respect to the amplitudes $\eta_1,\eta_2$.
Analytic solution of this system is possible only at negligibly small coefficient $|\gamma_2|$. This case at $\gamma_1>0$ we obtain
\begin{eqnarray}
\eta_2^2\cong -\frac{\alpha_{10}}{2\beta_1-\frac{\gamma_1^2}{8\beta M^2_0}}\left( T-T_{sc0}-\frac{\gamma_1M_0}{\alpha_{10}}  \right ),
\label{e}
\end{eqnarray}
\begin{eqnarray}
\eta_1\cong\frac{|\gamma_2|}{\alpha_1+\gamma_1M_0}~\eta_2.
\label{t}
\end{eqnarray}

This description of the  second order phase transition from ferromagnetic to the ferromagnetic superconducting state is valid
at the assumption $m<<M_0$. However, at pressure enhancement the Curie temperature and the critical temperature of superconducting transition (see Fig.1) approach each other, the value of $M_0$ gets smaller and according to Eq.(\ref{m}) the value of $m$ increases. One can expect  the turning of the second order transition into the first order transition
 such that the order parameters $\eta_1~,\eta_2,~m$ undergo  finite jumps from zero to the finite values at
temperature larger than the critical temperature given by Eq.(\ref{al}).  Indeed, the this type of behavior was established in Ref.9 by the numerical solution of  nonlinear equations for the order parameter components at close enough values $T_{Curie}$ and $T_{sc}$.

\subsection{Phase transitions from the normal to the ferromagnetic superconducting state}

To establish the whole phase diagram one must  consider the phase transition from the normal nonmagnetic state to the superconducting state. 
The  free energy density  (\ref{FE'}) minimization in respect $\eta_1,~\eta_2,~M$ yields
\begin{eqnarray}
\alpha_1\eta_1+\gamma_1M\eta_1-|\gamma_2|\eta_2
+2\beta_1\eta_1^3
+\beta_2\eta_1\eta_2^2=0,
\label{eta11}
\\
\alpha_1\eta_2-\gamma_1M\eta_2-|\gamma_2|\eta_1
+2\beta_1\eta_2^3+
\beta_2\eta_1^2\eta_2
=0,\label{eta22}
\\
2\alpha M+4\beta M^3+ \gamma_1(\eta_1^2-\eta_2^2)
=0.
\label{mo'}
\end{eqnarray}

At $\alpha>0$ there are two type  solutions of these equations such that
\begin {equation}
\eta_1=\eta_2,~~~~~~M=0,
\label{1}
\end{equation}
and
\begin {equation}
\eta_1\ne\eta_2,~~~~~M\ne 0.
\label{2}
\end{equation}

In the first case the transition to the ferromagnetic superconducting state occurs by means of  two consecutive phase transitions: the phase transition from the normal state to the nonmagnetic superconducting state
 followed at lower temperature  by the transition to the ferromagnetic superconducting state. 
In the second case the phase transition to the ferromagnetic superconducting state occurs directly from the normal state.
We consider  these situations separately.

\subsubsection{Two consecutive phase transitions from the normal to the ferromagnetic superconducting state}

The solution (\ref{1}) is realized at large enough positive $\alpha$ when formation of a ferromagnetic state is not
energetically profitable. This case 
the common magnitude of the superconducting amplitudes is
\begin{equation}
\eta^2=-\frac{\alpha_1-|\gamma_2|}{2\beta_1+\beta_2}.
\label{eta}
\end{equation}
At positive sum $2\beta_1+\beta_2>0$ this phase transition is of the second order and occurs at
\begin{equation}
T_{sc}=T_{sco}+\frac{|\gamma_2|}{\alpha_{10}},
\label{b}
\end{equation}
that coincides with Eq.(\ref{al}) at $M_0=0$.

To pass in the ferromagnetic superconducting state the system must undergo one more phase transition.
At this transition  the magnetization $M$ spontaneously appears  and the superconducting order parameter amplitudes acquire the deviations from the value given by Eq. (\ref{eta})
\begin{equation}
\eta_1=\eta +\delta_1,~~~~~ \eta_2=\eta+\delta_2.
\label{eta1eta2}
\end{equation}
The free energy acquires the following form
\begin{widetext}
\begin{eqnarray}
&F=\alpha M^2+\beta M^4+ \alpha_1(\delta_1^2+\delta_2^2)+\gamma_1M\left [2\eta(\delta_1-\delta_2)+\delta_1^2-\delta_2^2\right ] -2|\gamma_2|\delta_1\delta_2
\nonumber\\
&+\beta_1\left [ 6\eta^2(\delta_1^2+\delta_2^2)+4\eta(\delta_1^3+\delta_2^3)+\delta_1^4+\delta_2^4 \right ]
+\beta_2\left [\eta^2(\delta_1^2+\delta_2^2+4\delta_1\delta_2)+2\eta\delta_1\delta_2(\delta_1+\delta_2)+\delta_1^2\delta_2^2
\right ]
\end{eqnarray}
\end{widetext}
Here we have taken  into account that $\eta$
is the  minimum of free energy at $M=\delta_1= \delta_2=0$ and omitted the zero order 
 terms in respect of $M,\delta_1,\delta_2$ . The order parameters are determined from the conditions of the free energy minimum
\begin{equation}
\frac{\partial F}{\partial \delta_1}=0,~~~\frac{\partial F}{\partial \delta_2}=0,~~~\frac{\partial F}{\partial M}=0.
\end{equation}
One can easily check that in linear approximation the equations for $(\delta_1-\delta_2)$ and $M$
\begin{eqnarray}
&\left [ \alpha_1+|\gamma_2|+(6\beta_1-\beta_2)\eta^2 \right](\delta_1-\delta_2)+2\gamma_1\eta M=0,
\nonumber\\
&\gamma_1\eta(\delta_1-\delta_2)+\alpha M=0
\label{linear}
\end{eqnarray}
are decoupled from the equation for  $(\delta_1+\delta_2)$.  Hence, the latter combination is  of the next order of smallness  in comparison with 
\begin{equation}
M\propto (\delta_1-\delta_2)\propto\sqrt{T_{scM}-T}.
\end{equation}
 Here $T_{scM}$ is the critical temperature of transition from the superconducting to the superconducting ferromagnetic state which is determined  from the equation given by the equality to zero of  the  determinant of the system (\ref{linear})
\begin{equation}
\left [ \alpha_1+|\gamma_2|+(6\beta_1-\beta_2)\eta^2 \right]\alpha -2\left [\gamma_1\eta\right  ]^2=0.
\end{equation}

\subsubsection{Direct phase transition from the normal to the ferromagnetic superconducting state}

The second type solution (\ref{2}) is realized at small enough positive $\alpha$. The  analytical treatment
is possible
in neglect of  third order term $4\beta M^3$  in Eq.(\ref{mo'}), then
\begin{equation}
M\cong -\frac{\gamma_1(\eta_1^2-\eta_2^2)}{2\alpha}.
\label{Mo}
\end{equation}
Passing to the  sum and the difference of 
Eqs.(\ref{eta11}) and (\ref{eta22})   and using  the Eq. (\ref{Mo})  
we come to the equations
\begin{eqnarray}
\eta\left [\alpha_1-|\gamma_2|- 2\frac{\gamma_1^2}{\alpha}\delta^2 +2\beta_1(\eta^2+3\delta^2)+\beta_2(\eta^2-\delta^2)\right ]=0,~~\\
\delta\left [\alpha_1+|\gamma_2|- 2\frac{\gamma_1^2}{\alpha}\eta^2 +2\beta_1(3\eta^2+\delta^2)-\beta_2(\eta^2-\delta^2)\right ]=0,~~
\end{eqnarray}
where
\begin{equation}
\eta=\frac{1}{2}(\eta_1+\eta_2),~~~\delta=\frac{1}{2}(\eta_1-\eta_2).
\end{equation}
The solution of these equations at $\eta\ne 0,~\delta\ne 0$ is
\begin{eqnarray}
\eta^2=\frac{1}{2}\frac{\alpha_1\alpha}{\gamma_1^2-4\beta_1\alpha}+\frac{1}{2}\frac{|\gamma_2|\alpha}
{\gamma_1^2-(2\beta_1+\beta_2)\alpha},\\
\delta^2=\frac{1}{2}\frac{\alpha_1\alpha}{\gamma_1^2-4\beta_1\alpha}-\frac{1}{2}\frac{|\gamma_2|\alpha}
{\gamma_1^2-(2\beta_1+\beta_2)\alpha}.
\end{eqnarray}
Thus, at direct transition from the normal to the ferromagnetic superconducting state the  order parameter components  $M, \eta_1,\eta_2$ undergo the finite jumps. This is the phase transition of the first order. 

At a phase transition the free energy is not changed, that gives the equation for the phase transition temperature
\begin{eqnarray}
F=\alpha M^2+\beta M^4 + \alpha_1(\eta_1^2+\eta_2^2)+\gamma_1M(\eta_1^2-\eta_2^2)\nonumber\\-
2|\gamma_2|\eta_1\eta_2+
\beta_1(\eta_1^4+\eta_2^4)+\beta_2\eta_1^2\eta_2^2=0.~~~~~~~~~~~~~~
\label{0}
\end{eqnarray}
We solve this equation in the assumption that  the coefficient $|\gamma_2|$ is negligibly small. Then at small enough values of $\alpha$ one can use  the approximate  expressions
\begin{equation}
\eta^2\approx\delta^2\approx\frac{\alpha\alpha_1}{2\gamma_1^2},~~~~M\approx-2\frac{\gamma_1}{\alpha}\eta^2.
\end{equation}
Substituting  them to the  Eq.(\ref{0}) we obtain
\begin{equation}
\eta^4\left [\alpha_1-\frac{\gamma_1^2}{\alpha}+4\beta\right ]=0
\end{equation}
So, at small enough positive $\alpha$ the phase transition temperature is given by 
\begin{equation}
T_{sc}\approx T_{sco}+\frac{\gamma_1^2}{\alpha \alpha_{10}},
\label{bb}
\end{equation}
that  exceeds the critical temperature of the second order transition given by Eq. (\ref{b}).
 
 \subsection{Phase diagram}

The analytic derivation made at several not strongly restrictive assumptions leads to the conclusion   that the direct phase transition from the normal to the ferromagnetic superconducting state is of the first order.
This confirms the statement 
numerically established in Ref.9. On the other hand, as we have pointed out in the section {\bf IIIB.}, when the temperatures of phase transitions to the normal ferromagnetic and to the superconducting ferromagnetic state are close to each other,
  the transition from the ferromagnetic to the ferromagnetic superconducting state is of   the first order.

Thus, the simple phase diagram
with intersection of ferromagnetic and superconducting phase transition lines  like it is drown in Fig.1 cannot be realized at least in frame of the model under consideration.  Near the intersection of the critical lines of the ferromagnetic and the superconducting transitions there is a piece of the transition of the first order as this is
shown in Fig.2 by the thick line. The pressure interval where this transition takes place can be quite small. 
The presence of the first order transition  seems to be in correspondence with the sharp drop of resistivity at superconducting phase transition  in this pressure interval found in the paper \cite{Bastien16}.

\section{Magnetostatics }

The authors of Ref.9 discussed the phase transition between the SC state and FM+SC state in the neutral Fermi liquid.
 The situation is changed in a charged Fermi liquid because  the  magnetization  in the superconducting state is inevitably accompanied by the screening currents. The emergency of superconducting state in the ferromagnet  state of UCoGe takes place at finite magnetization.
 Whereas the arising of ferromagnetic state in the superconducting state of UCoGe  accompanied by the smooth increasing of magnetization from zero to the finite value. This determines the difference between the transition from the ferromagnetic to the superconducting ferromagnetic state and the transition from the superconducting to the superconducting ferromagnetic state which we discuss here.

Let us consider a cylindrical sample  of radius $R$  with axis parallel to the easy magnetization axis. 
A phase transition to the superconducting ferromagnetic state is accompanied by the appearance of super-currents. The  corresponding London equation for magnetic induction is
\begin{equation}
curl{\bf B}=\frac{4\pi{\bf j}}{c}=-\frac{{\bf A}}{\delta^2}+4\pi~curl{\bf M},
\end{equation}
where $\delta$ is the London penetration depth. The contribution to the current due to the term $c~curl{\bf M}$ is non-vanishing only in the surface layer with thickness of the order of coherence length $\xi<<\delta<<R$ \cite{Book}, hence, the  small enough magnetic field has  to decay in the sample volume:
\begin{equation}
{\bf B}( r )={\bf B}( R )\exp\left ( -\frac{R-r}{\delta} \right ),
\end{equation}
where the surface magnetic field is determined by the magnetic moment created by the super-currents flowing in the surface layer
\begin{equation}
{\bf B}( R )=4\pi{\bf M}.
\end{equation}
In UCoGe according to the phase diagram drown in Fig.2    the  formation of ferromagnetic superconducting state  occurs (i)  from the normal ferromagnetic state, (ii)  from the nonmagnetic  
normal state either through the two consecutive 
 phase transitions of the second order $N\to SC$ and then $SC\to FM+SC$  or directly by means of the first order transition.

\subsection{Magnetostatics below transition from the ferromagnetic to the ferromagnetic superconducting state} 

Just below the temperature of the phase transition  from the ferromagnetic to the ferromagnetic superconducting state  discussed in the section
 {\bf III.B} the field at surface is
\begin{equation}
{\bf B}( R )=4\pi\left(M_0 -\frac{\gamma_1(\eta_1^2-\eta_2^2)}{8\beta M_0^2}\right ).
\end{equation}
According to the  experimental results reported in \cite{Deguchi10,Paulsen12,Hykel14} this field at ambient pressure is larger than the lower critical field field $H_{c1}$
in UCoGe. So, the complete field screening is not realised, and  the phase transition occurs directly to the superconducting mixed state. The vortex cores occupy the small part of the sample volume, and almost whole volume is in the superconducting state with the order parameter given by Eqs.(\ref{e}), (\ref{t}). The specific heat jump
at phase transition to the superconducting state at the ambient pressure has the finite value
\begin{equation}
\Delta C\cong  \frac{\alpha_{10}T_{sc}}{2\beta_1-\frac{\gamma_1^2}{8\beta M^2_0}}.
\end{equation}
Here, we have neglected the temperature dependence of $M_0(T)$, just taking it as the constant $M_0=M_0(T_{sc})$.
The specific heat jump at this transition  has been registered \cite{Aoki14}.

\subsection{Magnetostatics below transition from the superconducting to the ferromagnetic superconducting state} 

Another situation takes place at the transition from the superconducting to the ferromagnetic superconducting state. This case  discussed in the section {\bf III.C} the surface field is determined by the small magnetic moment arising  at transition in magnetic superconducting state
\begin{equation}
{\bf B}( R )\cong4\pi M\propto \sqrt{T_{scM}-T}.
\end{equation}
This field is certainly smaller than the lower critical field in well developed superconducting state with superconducting density $n_s\propto \eta^2\propto (T_{sc}-T)$
\begin{equation}
H_{c1}=2\pi\mu_bn_s\ln\frac{\delta}{\xi}\propto (T_{sc}-T).
\end{equation}
The transition to the ferromagnetic superconducting state  is characterized by the emergency of magnetic moment $M$ and the magnetic part of superconducting ordering $\sim(\delta_1-\delta_2)$. But the super-currents completely  screen this magnetism in the bulk of material. 
The gradual increase of  magnetization from zero to some finite value is not accompanied  by a bulk phase transition as it is in the process of   the Meissner state formation in
 a  superconductor of the second kind 
 under external magnetic field smaller than $H_{c1}$.

The pressure decrease stimulates ferromagnetism.
Hence, at low temperatures and low pressures the magnetization will exceed the lower critical field and a sample passes to the ferromagnetic superconducting mixed state. So, instead  a phase transition between the superconducting and the ferromagnetic superconducting state one can expect just the transition between the Meissner  and the mixed superconducting states.

Thus, there  is no bulk phase transition at all. 
The   ferromagnetic  superconducting Meissner state exists in the region between two dashed lines 
shown in Fig.3. The  actual position of $H_{c1}$ line is subject to experimental dedetermination. It  can be in principle much more to the left, than it is drawn in Fig.3, and also as well as  more to the right  such that near 
 the first order transition the two dashed lines  can be merged to one line.

\section{Conclusion}

The Landau theory  allows to establish specific properties of phase transformations in anisotropic ferromagnetic superconducting material UCoGe.
There was found that the phase transition from the ferromagnetic to the ferromagnetic superconducting state at ambient pressure  is characterized by the appearance of superconducting part of the order parameter whereas  the ferromagnetic component
does undergo insignificant changes.  However,  at higher pressures this transition can turn to the transition of the first order.

There was proven that the direct phase transition from the nonmagnetic normal state to the ferromagnetic superconducting state is of the first order and exists in small pressure interval. Out of this interval the transition to the ferromagnetic superconducting state occurs by means of two consecutive phase transitions of the second order from normal to nonmagnetic superconducting state and then from  the nonmagnetic superconducting state to ferromagnetic superconducting state.
The model phase diagram  in neutral Fermi liquid acquires the  shape shown in 
 Fig.2. 

The phase diagram modification introduced by screening currents in charged Fermi liquid is shown in Fig.3.
The magnetic moment at phase transition from the ferromagnetic to the ferromagnetic superconducting state is just partially screened
by the superconducting currents. Whereas at phase transition from the superconducting state to the ferromagnetic superconducting state this screening is complete what shades the manifestations of a bulk phase  transition. The  ferromagnetic mixed  superconducting state occurs
only at lower pressures where the spontaneous magnetic moment exceeds the lower critical field. 
The   position of $H_{c1}$ line is subject to experimental determination.

\begin{figure}[p]
\includegraphics
[height=.8\textheight]
{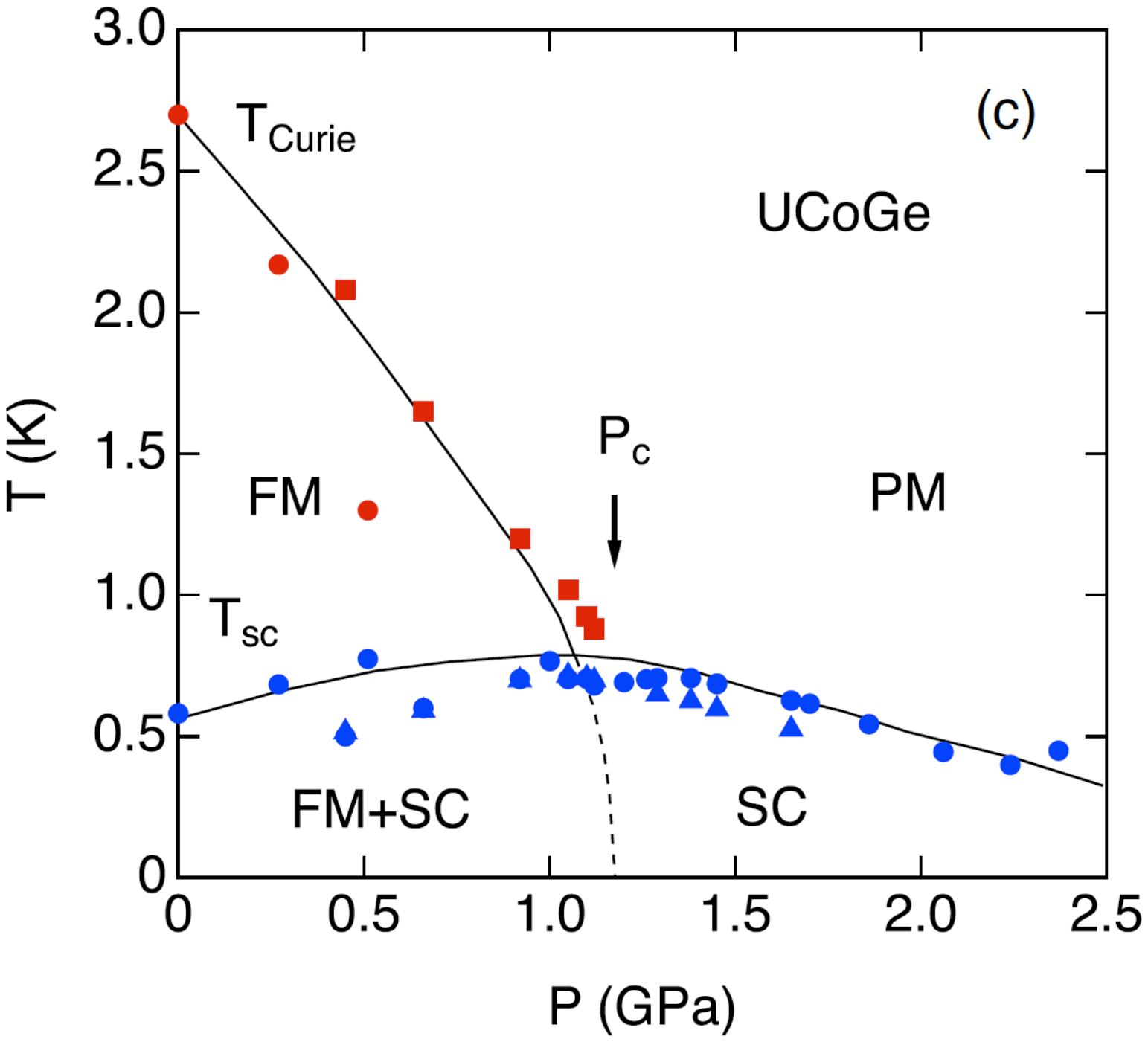}
 \caption{(Color online) Temperature-pressure phase diagram of 
 UCoGe. Notations  FM, SC and PM  used for ferromagnetic, superconducting and paramagnetic  phases correspondingly
 \cite{Aoki14}. }
\end{figure}

\begin{figure}[p]
\includegraphics
[height=.8\textheight]
{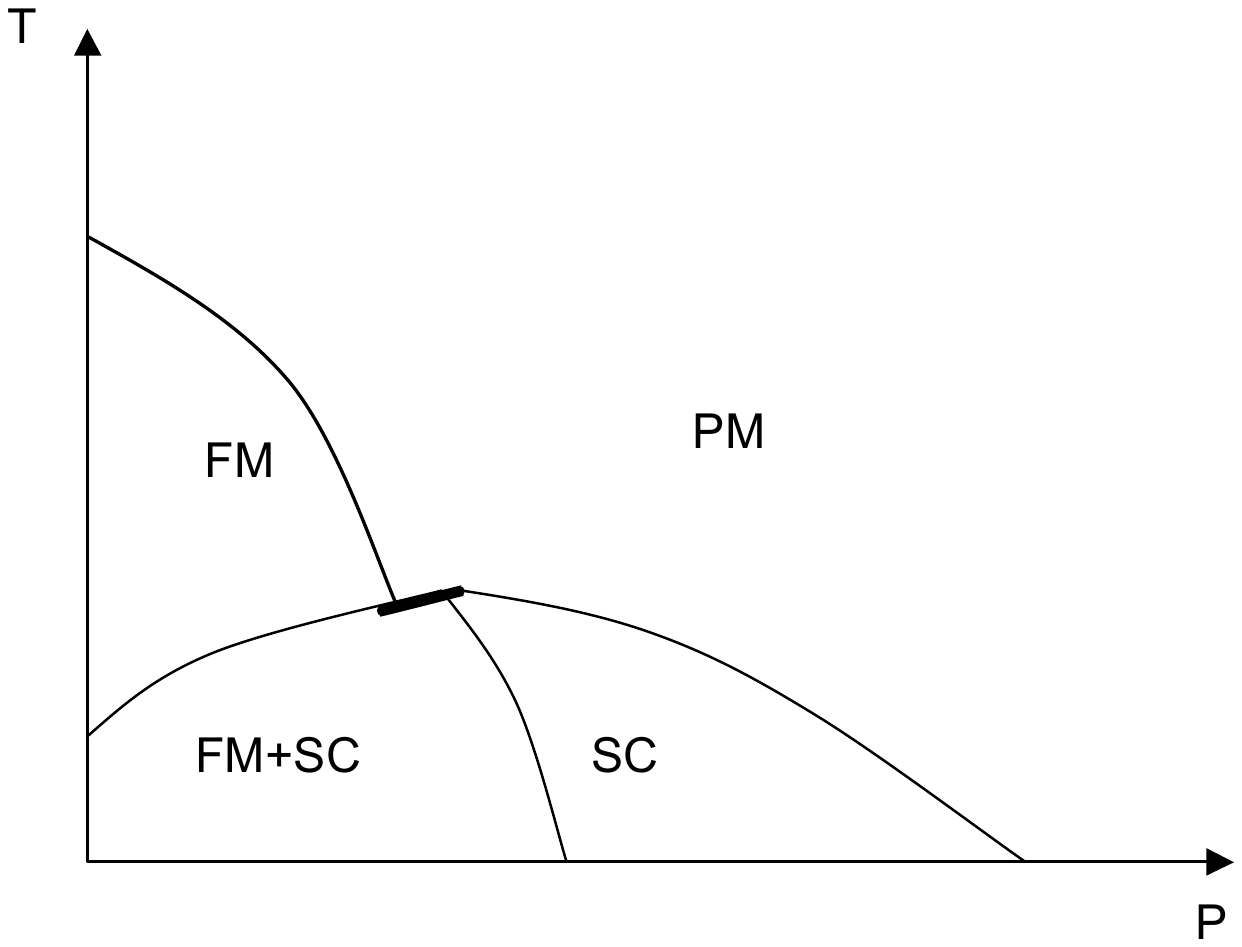}
 \caption{ Schematic temperature-pressure phase diagram of 
 UCoGe in frame of neutral Fermi liquid model. Notations  FM, SC and PM  used for ferromagnetic, superconducting and paramagnetic  phases correspondingly. The thin and thick lines are the  lines of the second and the first order transitions correspondingly. 
 }
\end{figure}

\begin{figure}[p]
\includegraphics
[height=.8\textheight]
{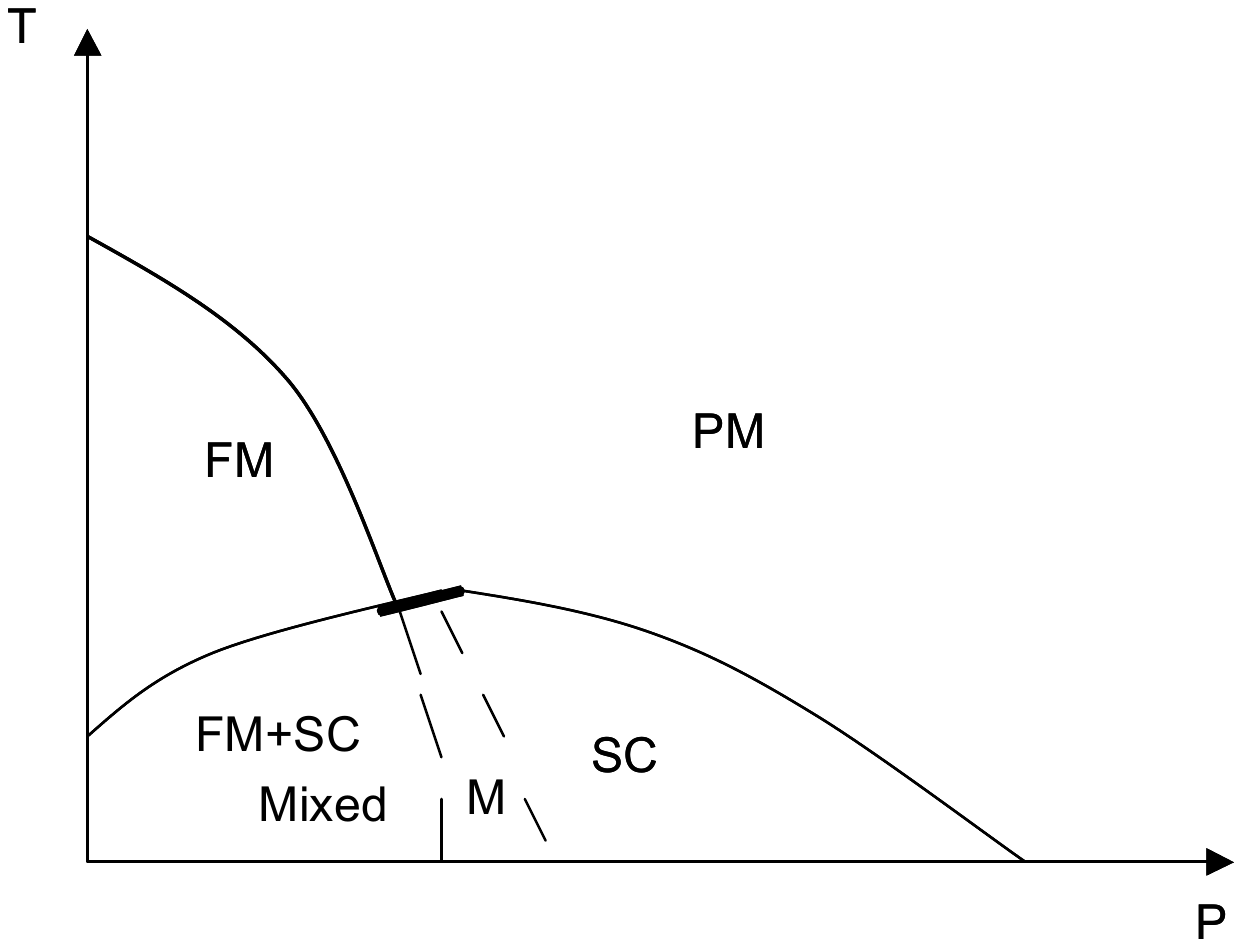}
 \caption{ Schematic temperature-pressure phase diagram of 
 UCoGe taking into account the effect of super-currents screening. Notations  FM, SC and PM  used for ferromagnetic, superconducting and paramagnetic  phases correspondingly. The thin and thick lines are the  lines of the second and the first order transitions correspondingly. The right dashed line is the imaginary line of the transition
 between the nonmagnetic and ferromagnetic Meissner superconducting states which is not  a phase transition in the bulk of sample. The left dashed line is the line of $H_{c1}$  dividing the Meissner and the mixed ferromagnetic superconducting states.}
\end{figure}

\end{document}